\documentstyle[12pt]{article}
\title{Relativistic Hartree-Bogoliubov description of the 
neutron drip-line in light nuclei}
\author{G.A. Lalazissis, D. Vretenar,
W. P\" oschl and P. Ring \\
Physik-Department der Technischen Universit\"at M\"unchen, \\
D-85747 Garching, Germany}
\begin{document}
\maketitle
\begin{abstract}
The Relativistic Hartree Bogoliubov theory is applied in
the mean-field approximation to the description of
properties of light nuclei with large neutron excess.
Pairing correlations and the coupling to particle continuum
states are described by finite range two-body forces. Using
standard parameter sets for the mean-field Lagrangian and
the pairing interaction of Gogny-type, self-consistent
solutions in coordinate space are calculated for the ground
states of a number of neutron-rich nuclei.  The model
predicts the location of the neutron drip-line, reduction
of the spin-orbit interaction, $rms$ radii, changes in
surface properties, the formation of neutron skin and of
neutron halo.
\end{abstract}
\newpage
\baselineskip = 24pt

%
\section {Introduction}

Experimental and theoretical studies of exotic nuclei with
extreme isospin values present one of the most active areas
of research in nuclear physics.  Experiments with
radioactive nuclear beams provide the opportunity to study
very short-lived nuclei with extreme neutron to proton
ratio N/Z. New theoretical models and techniques are being
developed in order to describe unique phenomena in nuclei
very different from those usually encountered along the
line of stability.  On the neutron-rich side in particular,
exotic phenomena include the weak binding of the outermost
neutrons, pronounced effects of coupling between bound
states and the particle continuum, regions of neutron halos
with very diffuse neutron densities and major modifications
if the shell structures. Neutron drip lines of relatively
light nuclei have become accessible in experiments, and
properties of these exotic objects are currently being
studied with a variety of theoretical approaches. Because
of their relevance to the r-process in nucleosynthesis,
nuclei close to the neutron drip line are also very
important in nuclear astrophysics.  Detailed knowledge of
their structure and properties would help the determination
of astrophysical conditions for the formation of neutron
rich stable isotopes.

The main problem in the theoretical description of
drip-line nuclei arises from the closeness of the Fermi
level to the particle continuum: particle-hole and pair
excitations reach the continuum. The coupling between bound
states and the particle continuum has to be explicitly
included in any model description.  The Relativistic
Hartree Bogoliubov (RHB) theory, which is a
relativistic extension of the Hartree Fock Bogoliubov
theory, provides a unified description of mean-field and
pairing correlations.  A fully self-consistent RHB theory
in coordinate space correctly describes the coupling
between bound and continuum states. The theory provides a
framework for describing the nuclear many-body problem as a
relativistic system of baryons and mesons. In the
self-consistent mean-field approximation, the theory has
been used to perform detailed calculations for a variety of
nuclear structure phenomena, as well as for the dynamics of
heavy-ion collisions. Although conventional
non-relativistic Hartree-Fock models and the relativistic
mean-field theory predict very similar properties for
nuclei close to the $\beta$-stability line, recent
applications to the structure of drip-line nuclei have
shown significant differences, especially in the isospin
dependence of the spin-orbit potential.

In Refs.~\cite{PVL.97,LVR.97} we have investigated, in the
framework of relativistic Hartree-Bogoliubov theory, light
nuclear systems with large neutron excess. The isospin
dependence of the spin-orbit interaction in light
neutron-rich nuclei has been studied in Ref.~\cite{LVR.97}.
In ~\cite{PVL.97} we have described the microscopic details
of the formation of neutron halo in the mass region above
the $s-d$ shell. Pairing correlations and the coupling to
particle continuum states have been described by finite
range two-body Gogny-type interaction. Finite element
methods have been used in the coordinate space
discretization of the coupled system of
Dirac-Hartree-Bogoliubov and Klein-Gordon equations
\cite{PVR1.97,PVR2.97}. Calculations have been performed
for the isotopic chains of Ne and C nuclei. We find
evidence for the occurrence of multi-neutron halo in
heavier Ne isotopes. The detailed microscopic analysis has
shown how the properties of the 1f-2p orbitals near the
Fermi level and the neutron pairing interaction determine
the formation of the halo. On the other hand our
calculations have shown that, although for Ne the neutron
drip was found at $N=32$, for C already the $N=20$ isotope
is unbound. Preliminary results for Oxygen have placed the
neutron drip at $N=18$ or $N=20$, depending on the
effective force used in the mean-field calculations. This
is a very interesting result: by adding just one or two
protons, the resulting mean-fields bind up to twelve more
neutrons.

In the present work we extend our investigation of
Ref.~\cite{PVL.97}, and study in detail the neutron
drip-line for the elements C, N, O, F, Ne, Na and Mg.  In
the region around $N=20$ we expect that these nuclei are to
a good approximation spherical, or that an eventual small
deformation would not change our results quantitatively. In
particular, we study how the proton number affects the
neutron single-particle levels around $N=20$ and the
position of the neutron drip. We also discuss changes in
the spin-orbit interaction, surface diffuseness and the
formation of the neutron halo. In Sec.~2 we present an
outline of the relativistic Hartree-Bogoliubov theory and
discuss the various approximations. In Sec.~3 the theory is
applied to the calculation of ground state properties of a
number of light neutron-rich nuclei. A summary of our
results is presented in Sec.~4.

\section {The relativistic Hartree-Bogoliubov model}
%

Detailed properties of nuclear matter and finite nuclei
along the $\beta$-stability line have been very
successfully described in the framework of relativistic
mean field models ~\cite{SW.86,Rei.89,Ser.92,Rin.96}.  In
comparison with conventional non-relativistic approaches,
relativistic models explicitly include mesonic degrees of
freedom and describe the nucleons as Dirac particles.
Nucleons interact in a relativistic covariant manner
through the exchange of virtual mesons:~the isoscalar
scalar $\sigma$-meson, the isoscalar vector $\omega$-meson
and the isovector vector $\rho$-meson.  The model is based
on the one boson exchange description of the
nucleon-nucleon interaction. We start from the effective
Lagrangian density
\begin{eqnarray}
{\cal L}&=&\bar\psi\left(i\gamma\cdot\partial-m\right)\psi
\nonumber\\
&&+\frac{1}{2}(\partial\sigma)^2-U(\sigma )
-\frac{1}{4}\Omega_{\mu\nu}\Omega^{\mu\nu}
+\frac{1}{2}m^2_\omega\omega^2
-\frac{1}{4}{\vec{\rm R}}_{\mu\nu}{\vec{\rm R}}^{\mu\nu}
+\frac{1}{2}m^2_\rho\vec\rho^{\,2}
-\frac{1}{4}{\rm F}_{\mu\nu}{\rm F}^{\mu\nu}
\nonumber\\
&&-g_\sigma\bar\psi\sigma\psi-
g_\omega\bar\psi\gamma\cdot\omega\psi-
g_\rho\bar\psi\gamma\cdot\vec\rho\vec\tau\psi -
e\bar\psi\gamma\cdot A \frac{(1-\tau_3)}{2}\psi\;.
\label{lagrangian}
\end{eqnarray}
Vectors in isospin space are denoted by arrows, and
bold-faced symbols will indicate vectors in ordinary
three-dimensional space.  The Dirac spinor $\psi$ denotes
the nucleon with mass $m$.  $m_\sigma$, $m_\omega$, and
$m_\rho$ are the masses of the $\sigma$-meson, the
$\omega$-meson, and the $\rho$-meson.  $g_\sigma$,
$g_\omega$, and $g_\rho$ are the corresponding coupling
constants for the mesons to the nucleon. $e^2 /4 \pi =
1/137.036$.  Coupling constants and unknown meson masses
are parameters, adjusted to fit data on nuclear matter and
finite nuclei.  $U(\sigma)$ denotes the non-linear $\sigma$
self-interaction \cite{BB.77}
\begin{equation}
U(\sigma)~=~\frac{1}{2}m^2_\sigma\sigma^2+\frac{1}{3}g_2\sigma^3+
\frac{1}{4}g_3\sigma^4,
\label{NL}
\end{equation}
and $\Omega^{\mu\nu}$, $\vec R^{\mu\nu}$, and $F^{\mu\nu}$
are field tensors
\begin{eqnarray}
\Omega^{\mu \nu} & = & \partial^{\mu} \omega^{\nu} - 
                       \partial^{\nu} \omega^{\mu} \\
\vec{R}^{\mu \nu} & = & \partial^{\mu} \vec{\rho}^{\,\nu} -
                        \partial^{\nu} \vec{\rho}^{\,\mu}\\
F^{\mu \nu} & = & \partial^{\mu} A^{\nu} - 
                       \partial^{\nu} A^{\mu}.
\end{eqnarray}
The lowest order of the quantum field theory is the {\it
mean-field} approximation: the meson field operators are
replaced by their expectation values.  The $A$ nucleons,
described by a Slater determinant $|\Phi\rangle$ of
single-particle spinors $\psi_i,~(i=1,2,...,A)$, move
independently in the classical meson fields.  The sources
of the meson fields are defined by the nucleon densities
and currents.  The ground state of a nucleus is described
by the stationary self-consistent solution of the coupled
system of Dirac and Klein-Gordon equations.  In the static
case for an even-even system, there will be no currents in
the nucleus, and the spatial vector components
\mbox{\boldmath $\omega,~\rho_3$} and ${\bf  A}$ vanish.
The Dirac equation reads
\begin{equation}
\label{statDirac}
\left\{-i\mbox{\boldmath $\alpha$}
\cdot\mbox{\boldmath $\nabla$}
+\beta(m+g_\sigma \sigma)
+g_\omega \omega^0+g_\rho\tau_3\rho^0_3
+e\frac{(1-\tau_3)}{2} A^0\right\}\psi_i=
\varepsilon_i\psi_i
\end{equation}
The effective mass $m^*({\bf r})$ is defined
\begin{equation}
\label{effmass}
m^*({\bf r})=m+g_{\sigma}\,\sigma({\bf r}),
\end{equation}
and the potential $V({\bf r})$ 
\begin{equation}
\label{scapot}
V({\bf r})=g_{\omega}\,\omega^0({\bf r})+
g_{\rho}\,\tau_3\,\rho^0_3({\bf r})
+e{{(1-\tau_3)}\over 2}A^0({\bf r}).
\end{equation}
In order to describe ground-state properties of spherical
open-shell nuclei, pairing correlations have to be taken
into account. For nuclei close to the $\beta$-stability
line, pairing has been included in the relativistic
mean-field model in the form of a simple BCS
approximation~\cite{GRT.90}.  However, for nuclei far from
stability the BCS model presents only a poor approximation.
In particular, in drip-line nuclei the Fermi level is found
close to the particle continuum.  The lowest particle-hole
or particle-particle modes are often embedded in the
continuum, and the coupling between bound and continuum
states has to be taken into account explicitly.  The BCS
model does not provide a correct description of the
scattering of nucleonic pairs from bound states to the
positive energy continuum~\cite{DFT.84,DNW.96}. It leads to
an unbound system, because levels in the continuum are
partially occupied. Including the system in a box of finite
size leads to unreliable predictions for nuclear radii
depending on the size of this box. In the non-relativistic
case, a unified description of mean-field and pairing
correlations is obtained in the  framework of the
Hartree-Fock-Bogoliubov (HFB) theory in coordinate
space~\cite{DFT.84}. The ground state of a nucleus $\vert
\Phi >$ is represented as the vacuum with respect to
independent quasi-particles. The quasi-particle operators
are defined by a unitary Bogoliubov transformation of the
single-nucleon creation and annihilation operators.  The
generalized single-particle Hamiltonian of HFB theory
contains two average potentials: the self-consistent field
$\hat\Gamma$ which encloses all the long range {\it ph}
correlations, and a pairing field $\hat\Delta$ which sums
up the {\it pp}-correlations. The expectation value of the
nuclear Hamiltonian $< \Phi\vert \hat H \vert \Phi >$ can
be expressed as a function of the hermitian density matrix
$\rho$, and the antisymmetric pairing tensor $\kappa$. The
variation of the energy functional with respect to $\rho$
and $\kappa$ produces the single quasi-particle
Hartree-Fock-Bogoliubov equations~\cite{RS.80}
\begin{eqnarray}
\label{equ.2.1}
\left( \matrix{ \hat h - \lambda & \hat\Delta \cr
                -\hat\Delta^* & -\hat h +\lambda
                 } \right) \left( \matrix{ U_k \cr V_k}\right) =
E_k\left( \matrix{ U_k \cr V_k } \right).
\end{eqnarray}
HFB-theory, being a variational approximation, results in a
violation of basic symmetries of the nuclear system, among
which the most important is the non conservation of the
number of particles.  In order that the expectation value
of the particle number operator in the ground state equals
the number of nucleons, equations (\ref{equ.2.1}) contain a
chemical potential $\lambda$ which has to be determined by
the particle number subsidiary condition. The column
vectors denote the quasi-particle wave functions, and $E_k$
are the quasi-particle energies.

If the pairing field $\hat \Delta$ is a constant,
HFB reduces to the BCS-approximation. The lower and upper
components $U_k(r)$ and $V_k(r)$ are proportional, with the
BCS-occupation amplitudes as proportionality constants.
For a more general pairing
interaction this will no longer be the case. As opposed to the
functions $U_k(r)$, the lower components $V_k(r)$ 
are localized functions of {\bf r}, as long as the
chemical potential $\lambda$ is below the continuum limit.
Since the densities are bilinear products of $V_k(r)$
(see Eqs.(\ref{equ.2.3.e})-(\ref{equ.2.3.h})), the system
is always localized. According to the theorem of Bloch and
Messiah \cite{RS.80}, an HFB wave function can be written
either in the quasiparticle basis as a product of
independent quasi-particle states, or in the {\it canonical basis}
as a highly correlated BCS-state. In the {\it canonical basis}
nucleons occupy  single-particle states. 
If the chemical potential is close to the
continuum, the pairing interaction scatters pairs of nucleons 
into continuum states. Because of additional pairing correlations,
particles which occupy those levels cannot evaporate. Mathematically
this is expressed by an additional
non-trivial transformation which connects the particle
operators in the canonical basis to the quasi-particles of
the HFB wave function.

The relativistic extension of the HFB theory was introduced
in Ref.~\cite{KR1.91}. In the Hartree approximation for
the self-consistent mean field, the Relativistic
Hartree-Bogoliubov (RHB) equations read
\begin{eqnarray}
\label{equ.2.2}
\left( \matrix{ \hat h_D -m- \lambda & \hat\Delta \cr
                -\hat\Delta^* & -\hat h_D + m +\lambda} \right) 
\left( \matrix{ U_k({\bf r}) \cr V_k({\bf r}) } \right) =
E_k\left( \matrix{ U_k({\bf r}) \cr V_k({\bf r}) } \right).
\end{eqnarray}
where $\hat h_D$ is the single-nucleon Dirac
Hamiltonian(\ref{statDirac}), and $m$ is the nucleon mass.
The RHB equations have to be solved self-consistently, with
potentials determined in the mean-field approximation from
solutions of Klein-Gordon equations
\begin{eqnarray}
\label{equ.2.3.a}
\bigl[-\Delta + m_{\sigma}^2\bigr]\,\sigma({\bf r})&=&
-g_{\sigma}\,\rho_s({\bf r})
-g_2\,\sigma^2({\bf r})-g_3\,\sigma^3({\bf r})   \\
\label{equ.2.3.b}
\bigl[-\Delta + m_{\omega}^2\bigr]\,\omega^0({\bf r})&=&
-g_{\omega}\,\rho_v({\bf r}) \\
\label{equ.2.3.c}
\bigl[-\Delta + m_{\rho}^2\bigr]\,\rho^0({\bf r})&=&
-g_{\rho}\,\rho_3({\bf r}) \\
\label{equ.2.3.d}
-\Delta \, A^0({\bf r})&=&e\,\rho_p({\bf r}).
\end{eqnarray}
for the sigma meson, omega meson, rho meson and photon
field, respectively.  The spatial components $\bf\omega$,
$\bf\rho$, and {\bf A} vanish due to time reversal
symmetry. Because of charge conservation, only the
3-component of the isovector rho meson contributes. The
source terms in equations (\ref{equ.2.3.a}) to
(\ref{equ.2.3.d}) are sums of bilinear products of baryon
amplitudes
\begin{eqnarray}
\label{equ.2.3.e}
\rho_s({\bf r})&=&\sum\limits_{E_k > 0} 
V_k^{\dagger}({\bf r})\gamma^0 V_k({\bf r}), \\
\label{equ.2.3.f}
\rho_v({\bf r})&=&\sum\limits_{E_k > 0} 
V_k^{\dagger}({\bf r}) V_k({\bf r}), \\
\label{equ.2.3.g}
\rho_3({\bf r})&=&\sum\limits_{E_k > 0} 
V_k^{\dagger}({\bf r})\tau_3 V_k({\bf r}), \\
\label{equ.2.3.h}
\rho_{\rm em}({\bf r})&=&\sum\limits_{E_k > 0} 
V_k^{\dagger}({\bf r}) {{1-\tau_3}\over 2} V_k({\bf r}).
\end{eqnarray}
where the sums run over all positive energy states. For M
degrees of freedom, for example number of nodes on a radial
mesh, the HB equations are 2M-dimensional and have 2M
eigenvalues and eigenvectors. To each eigenvector $(U_k,
V_k)$ with eigenvalue $E_k$, there corresponds an
eigenvector $(V^*_k, U^*_k)$ with eigenvalue $-E_k$. Since
baryon quasi-particle operators satisfy fermion commutation
relations, it is forbidden to occupy the levels $E_k$ and
$-E_k$ simultaneously. Usually one chooses the M positive
eigenvalues $E_k$ for the solution that corresponds to a
ground state of a nucleus with even particle number.

The system of equations (\ref{equ.2.2}), and
(\ref{equ.2.3.a}) to (\ref{equ.2.3.d}), is solved
self-consistently in coordinate space by discretization on
the finite element mesh~\cite{PVR1.97,PVR2.97}.  In the
coordinate space representation of the pairing field
$\hat\Delta $ in (\ref{equ.2.2}), the kernel of the
integral operator is
\begin{equation}
\label{equ.2.5}
\Delta_{ab} ({\bf r}, {\bf r}') = {1\over 2}\sum\limits_{c,d}
V_{abcd}({\bf r},{\bf r}') {\bf\kappa}_{cd}({\bf r},{\bf r}').
\end{equation}
where $a,b,c,d$ denote all quantum numbers, apart from the
coordinate $\bf r$, that specify the single-nucleon states.
$V_{abcd}({\bf r},{\bf r}')$ are matrix elements of a
general two-body pairing interaction, and the pairing
tensor is defined as
\begin{equation}
{\bf\kappa}_{cd}({\bf r},{\bf r}') = 
\sum_{E_k>0} U_{ck}^*({\bf r})V_{dk}({\bf r}').
\end{equation}
The integral operator $\hat\Delta$ acts on the wave function
$V_k({\bf r})$:
\begin{equation}
\label{equ.2.4}
(\hat\Delta V_k)({\bf r}) 
= \sum_b \int d^3r' \Delta_{ab} ({\bf r},{\bf r}') V_{bk}({\bf r}'). 
\end{equation}

The eigensolutions of Eq. (\ref{equ.2.2}) form a set of
orthogonal (normalized) single quasi-particle states. The corresponding
eigenvalues are the single quasi-particle energies.
In the self-consistent iteration procedure we work
in the basis of quasi-particle states. The self-consistent quasi-particle
eigenspectrum is then transformed into the canonical basis of single-particle
states. The canonical basis is defined to be the one
in which the matrix
$R_{kk'}=\bigl< V_k({\bf r})\big\vert V_{k'}({\bf r})\bigl>$ is diagonal.
The transformation to the canonical basis determines the energies
and occupation probabilities of single-particle states, that correspond
to the self-consistent solution for the ground state of a nucleus.

For nuclear systems with spherical symmetry the fields
$\sigma(r),\,\omega^0(r),\,\rho^0(r),$ and $A^0(r)$ depend
only on the radial coordinate $r$.  The nucleon spinors
$U_k$ ($V_k$) in (\ref{equ.2.2}) are characterized by the
angular momentum $j$, its $z$-projection $m$, parity $\pi$
and the isospin $t_3=\pm {1\over 2}$ for neutron and
proton. The two Dirac spinors $U_k({\bf r})$ and $V_k({\bf
r})$ form a {\it super-spinor}
\begin{equation}
\Phi_k({\bf r})~=~\pmatrix{ U_k({\bf r}) \cr V_k({\bf r})\cr}
\end{equation}
where
\begin{eqnarray}
\label{spherspinor}
{U_k(V_k)}({\bf r},s,t_3)=
\pmatrix{ g_{U(V)}(r)\Omega_{j,l,m} (\theta,\varphi,s) \cr
        if_{U(V)}(r)\Omega_{j,\tilde l,m} (\theta,\varphi,s) \cr }
         \chi_\tau(t_{3}).
\end{eqnarray}
$g(r)$ and $f(r)$ are radial amplitudes, $\chi_\tau$ is the
isospin function, the orbital angular momenta $l$ and
$\tilde l$ are determined by $j$ and the parity $\pi$
\begin{eqnarray}
l=\left\{ \matrix{ j+1/2 & \mbox{ for} & \pi = (-1)^{j+1/2} \cr
                  j-1/2 & \mbox{ for} & \pi = (-1)^{j-1/2} \cr } \right., 
\end{eqnarray}
and
\begin{eqnarray}
\tilde l=\left\{ \matrix{ j-1/2 & \mbox{ for} & \pi = (-1)^{j+1/2} \cr
                   j+1/2 & \mbox{ for} & \pi = (-1)^{j-1/2} \cr } \right. 
\end{eqnarray}
$\Omega_{jlm}$ is the tensor product of the orbital and
spin functions
\begin{equation}
\Omega_{j,l,m} (\theta,\varphi,s)=\sum\limits_{m_s,m_l}
\bigl< {1\over 2}m_slm_l\big\vert jm\bigr> 
\chi_{{1\over 2} m_s} Y_{lm_l}(\theta,\varphi).
\end{equation}
It is useful to define a single angular quantum number
$\kappa$ as the eigenvalue of the operator
$(1+\hat{\bf\sigma}\cdot{\hat{\bf l}}\,)$
\begin{equation}
\label{equ.2.11}
(1+\hat{\bf\sigma}\cdot{\hat{\bf l}}\,)\,\Omega_{\kappa,m} 
= -\kappa\,\Omega_{\kappa,m},
\end{equation}
\begin{equation}
\kappa=\pm(j+{1/2})~~~~ {\rm for}~~~~ j=l\mp{1/2}.
\end{equation}
$\kappa = \pm 1,\pm 2,\pm 3,...$.  The equations for the
radial amplitudes $g_{U(V)}(r)$ and $f_{U(V)}(r)$ are
derived from Eq. (\ref{equ.2.2}). The radial
single-particle Dirac Hamiltonian reads
\begin{equation}
\label{equ.2.13}
\hat h_{D}(r)=
\sigma_3\sigma_1 (\partial_r+r^{-1})-\sigma_1\kappa r^{-1}+
\sigma_3 m+\sigma_3 S(r) + V_0(r),
\end{equation}
where the scalar and vector potentials are
\begin{equation}
S(r) = g_\sigma \sigma(r),~~~ {\rm and}~~~
V^0(r)=g_{\omega}\,\omega^0(r)+
g_{\rho}\,\tau_3\,\rho^0_3(r)
+e{{(1-\tau_3)}\over 2}A^0(r).
\end{equation}
$\sigma_i$ (i=1,2,3) are the Pauli matrices. The integral
operator of the pairing interaction takes the form
\begin{equation}
\label{equ.2.14}
\hat\Delta (r) =\int\limits_0^{\infty}dr'{r'}^2 \Delta(r,r') 
\end{equation}
The kernel of the integral operator is defined 
\begin{equation}
\label{equ.2.15}
\Delta_{aa'}^{JM}(r,r')={1\over 2}\sum_{\tilde a,\tilde a'}
\bigl< r a,r' a'\big\vert
\hat V\big\vert r\tilde a,r'\tilde a'\bigr>_{JM}
{\bf\kappa}_{\tilde a,\tilde a'}(r,r')
\end{equation}
where $r$ and $r'$ denote radial coordinates, $a$, $a'$,
$\tilde a$ and $\tilde a'$ are quantum numbers that
completely specify single-nucleon states: $(n,l,j,m)$ or
$(n,\kappa,m)$, $J$ and $M$ are the total angular momentum
of the pair, and its $z$-projection, respectively.  In the
particle-particle ($pp$) channel the pairing interaction is
approximated by a two-body finite range Gogny interaction
\begin{equation}
V^{pp}(1,2)~=~\sum_{i=1,2}
e^{-(( {\bf r}_1- {\bf r}_2)
/ {\mu_i} )^2}\,
(W_i~+~B_i P^\sigma 
-H_i P^\tau -
M_i P^\sigma P^\tau),
\end{equation}
with parameters $\mu_i$, $W_i$, $B_i$, $H_i$ and $M_i$
$(i=1,2)$.  For the $J=0$ contribution to the pairing
matrix elements, the kernel of the integral operator reads
\begin{equation}
\label{equ.2.20}
\Delta_{\kappa} (r,r') = {1\over 2}\sum\limits_{\tilde\kappa}
V^{J=0}_{\kappa\tilde\kappa}(r,r') {\bf\kappa}_{\tilde\kappa}( r, r').
\end{equation}
The pairing tensor is calculated 
\begin{equation}
\label{equ.2.21}
{\bf\kappa}_{\kappa}(r,r')=\sum\limits_n 2\vert\kappa\vert
\pmatrix{
g^{(U)}_{n,\kappa}(r)g_{n,\kappa}^{(V)}(r') & 0 \cr
  0 & f^{(U)}_{n,\kappa}(r)f_{n,\kappa}^{(V)}(r') }
\end{equation}
where, for a quantum number $\kappa$,  the sum runs over
all solutions in the specified energy interval $0 < E <
E_{\rm max}$.  If we define the Dirac spinors
\begin{equation}
\Phi_U(r):=\pmatrix{g_U(r)\cr i f_U(r)}~~~~~~{\rm and}~~~~~~ 
\Phi_V(r):=\pmatrix{g_V(r)\cr i f_V(r)} ,
\end{equation}
the radial Dirac-Hartree-Bogoliubov equations read
\begin{eqnarray}
\label{equ..2.22}
(\hat h_D(r)-m -\lambda )\Phi_U(r)+
\int_0^{\infty}dr'r'^2\Delta(r,r')\Phi_V(r') 
= E\Phi_U(r) \nonumber \\
(-\hat h_D(r)+m +\lambda )\Phi_V(r)+
\int_0^{\infty}dr'r'^2\Delta(r,r')\Phi_U(r') 
= E\Phi_V(r)
\end{eqnarray}
The meson and photon fields are solution of the
Klein-Gordon equations
\begin{eqnarray}
\label{equ.2.23.a}
\bigl(-\partial_r^2 - {2\over r}\partial_r 
+ m_{\sigma}^2\bigr)\,\sigma(r)&=&-g_{\sigma}\,\rho_s(r)
-g_2\,\sigma^2(r)-g_3\,\sigma^3(r)   \\
\label{equ.2.23.b}
\bigl(-\partial_r^2 - {2\over r}\partial_r 
+ m_{\omega}^2\bigr)\,\omega^0(r)&=&
-g_{\omega}\,\rho_v(r) \\
\label{equ.2.23.c}
\bigl(-\partial_r^2 - {2\over r}\partial_r 
+ m_{\rho}^2\bigr)\,\rho^0(r)&=&-g_{\rho}\,\rho_3(r) \\
\label{equ.2.23.d}
\bigl(-\partial_r^2 - {2\over r}\partial_r \bigr)
\, A^0(r)&=& e\,\rho_p(r).
\end{eqnarray}
%
%
%
\section {Neutron drip-line in light nuclei}

The details of the neutron drip-line will depend on the
single-particle levels, position of the Fermi energy,
pairing interaction and  coupling between bound and
continuum states. In order to ascertain that our
microscopic predictions do not crucially depend on the
choice of the effective interaction,  we perform
calculations for three parameter sets of the mean-field
Lagrangian: NL1~\cite{RRM.86}, NL3~\cite{LKR.97}, and
NL-SH~\cite{SNR.93}.  The effective forces NL1 and NL-SH
have been frequently used to calculate properties of
nuclear matter and of finite nuclei, and have become
standard parameterizations for relativistic mean-field
calculations.  The parameter set NL3 has been derived
recently~\cite{LKR.97} by fitting ground state properties
of a large number of spherical nuclei.  Properties
calculated with the NL3 effective interaction are found to
be in very good agreement with experimental data for nuclei
at and away from the line of $\beta$-stability.

In the pairing channel one should in principle use a
one-meson exchange interaction $V_{abcd}$ derived by the
elimination of the mesonic degrees of freedom in the
Lagrangian (\ref{lagrangian}).  However, as it was shown in
Ref.~\cite{KR1.91}, the standard parameter sets of the
mean-field approximation lead to completely unrealistic
pairing matrix elements.  These parameters do not reproduce
the scattering data.  Since at present a relativistic
pairing force with parameters consistent with those of the
mean-field potential is not available, we use a
phenomenological Gogny-type finite range interaction in the
$pp$-channel, a procedure which requires no cut-off and
which provides a very reliable description of pairing
properties in finite nuclei. The parameter set
D1S~\cite{BGG.84} is used for the finite range Gogny type
interaction.

In Ref.~\cite{PVL.97} we have studied in detail the
formation of the neutron halo in Ne isotopes. In order to 
be more complete and as an introduction to the discussion
that follows, we briefly review the main arguments and
results of that work.  In Fig. 1a the $rms$ radii for Ne
isotopes are plotted as functions of neutron number. We
display neutron and proton $rms$ radii, and the N$^{1/3}$
curve normalized so that it coincides with the neutron
radius in $^{20}$Ne. The neutron $rms$ radii follow the
mean-field N$^{1/3}$ curve up to N $\approx$ 22.  For
larger values of N the neutron radii display a sharp
increase, while the proton radii stay practically constant.
This sudden increase in neutron $rms$ radii has been
interpreted as evidence for the formation of a
multi-particle halo. The phenomenon was also observed in
the plot of proton and neutron density distributions.  The
proton density profiles do not change with the number of
neutrons, while the neutron density distributions display
an abrupt change between $^{30}$Ne and $^{32}$Ne.  The
microscopic origin of the neutron halo has been found in a
delicate balance of the self-consistent mean-field and the
pairing field. This is shown in Fig. 1b, where we display
the neutron single-particle states 1f$_{7/2}$, 2p$_{3/2}$
and  2p$_{1/2}$ in the canonical basis, and the Fermi
energy as function of the neutron number.  For N$\leq$ 22
the triplet of states is high in the continuum, and the
Fermi level uniformly increases toward zero. The triplet
approaches zero energy, and a gap is formed between these
states and all other states in the continuum. The shell
structure dramatically changes at N$\geq$ 22. Between N =
22 and N = 32 the Fermi level is practically constant and
very close to the continuum. The addition of neutrons in
this region of the drip does not increase the binding. Only
the spatial extension of neutron distribution displays an
increase.  The formation of the neutron halo is related to
the quasi-degeneracy of the triplet of states 1f$_{7/2}$,
2p$_{3/2}$ and  2p$_{1/2}$. The pairing interaction
promotes neutrons from the 1f$_{7/2}$ orbital to the 2p
levels. Since these levels are so close in energy, the
total binding energy does not change significantly. Due to
their small centrifugal barrier, the 2p$_{3/2}$ and
2p$_{1/2}$ orbitals form the halo. The last bound isotope
is $^{40}$Ne. For N $\geq$ 32 (A$\geq$42) the neutron Fermi
level becomes positive, and heavier isotopes are not bound
any more. A similar mechanism has been studied in 
Ref.~\cite{MR.96}, for the experimentally observed halo in 
the nucleus $^{11}$Li. There the formation of the halo
is determined by the pair of neutron levels 1p$_{1/2}$ and 
2s$_{1/2}$. A giant halo has been also predicted for 
Zirconium isotopes~\cite{MR.97}. In that case the 
halo originates from the neutron orbitals
2f$_{7/2}$, 3p$_{3/2}$ and 3p$_{1/2}$.

In Figs. 2, 3 and 4 we display the neutron single-particle
energies in canonical basis for C, N, O, F, Ne, Na and Mg,
in the region around neutron number N=20. In Figs. 2 and 3
the left hand side columns display results calculated with
the NL1 effective interaction, NL3 results are shown in the
central panels, and on the right hand side levels that
correspond to the NL-SH parameterization are shown, as
function of neutron number. For Carbon, in Fig. 4 we
display results obtained with the NL3 and NL-SH parameter
sets.  The NL1 force does not produce a self-consistent
solution for the ground states of C isotopes. We display
the energy levels of the triplet of neutron states
1f$_{7/2}$, 2p$_{3/2}$ and  2p$_{1/2}$, the neutron Fermi
level and, as reference, the occupied orbital 1d$_{3/2}$.
Other states in the continuum, as well as deep hole states,
are too far away from the Fermi level and therefore not
important in the present consideration. Results calculated
with different effective interactions are rather similar,
and in agreement with our conclusions for Ne isotopes: the
location of the neutron drip is essentially determined by
the triplet 1f$_{7/2}$, 2p$_{3/2}$ and  2p$_{1/2}$.  It
appears that NL1 leads to strongest binding, followed by
NL3 and NL-SH. For example, $^{28}$O is bound for NL1 and
NL3, but unbound for the NL-SH parameter set.  As the
neutron number N=20 is approached, for C, N, O and F the
triplet is still high in the continuum, and the pairing
interaction does not have enough strength to promote
neutrons into these levels. Depending on the effective
force, the neutron drip is found at N=18 or N=20. Starting
from Ne, as we have already shown in Fig. 1, the triplet
1f$_{7/2}$, 2p$_{3/2}$ and  2p$_{1/2}$ is much lower and
neutrons above N=20 begin to fill the 1f$_{7/2}$ orbital.
For Ne and Na, 2p$_{3/2}$ and  2p$_{1/2}$ are close to the
1f$_{7/2}$, and therefore they are also populated forming
the neutron skin, or eventually the neutron halo. For Mg
the neutrons above the $s-d$ shell predominantly populate
the 1f$_{7/2}$ orbital, and one expects that deformation
plays an important role in the precise location of the
neutron drip.

An interesting result is observed if one plots the neutron
single-particle levels as function of the number of
protons. This is done in Fig. 5 for N=18, N=20 and N=22
isotones. The levels 2p$_{3/2}$, 2p$_{1/2}$, 1f$_{7/2}$,
and 1f$_{5/2}$, as well as the neutron Fermi level, are
displayed for the three effective forces. For a given
number of neutrons, increasing the proton number means
going toward the $\beta$-stability, i.e. away from the
neutron drip-line. The binding increases, as reflected in
the negative slope of the Fermi level.  What we find very
interesting is that the line that corresponds to the Fermi
level is almost parallel to the 1f$_{7/2}$ level. It
appears that the binding of neutrons in these systems is
determined by the position of the 1f$_{7/2}$ orbital. The
levels 2p$_{3/2}$, 2p$_{1/2}$, and 1f$_{5/2}$ change very
little with Z. This is also true for 1g$_{9/2}$, not shown
in the figure, at an energy of $\approx$ 20 MeV. On the
other hand, the energy of the 1f$_{7/2}$ orbital decreases
considerably by adding more protons. If the spin-orbit
splittings are compared, we notice that the energy
difference for the doublet 2p$_{3/2}$ - 2p$_{1/2}$ is
almost constant as function of Z, while a very strong
isospin dependence is observed for 1f$_{7/2}$ - 1f$_{5/2}$.
This is the crucial point and requires a careful
explanation.  In Ref.~\cite{LVR.97} we have analyzed the
isospin dependence of the spin-orbit interaction for Ne and
Mg isotopes. We have shown that the corresponding term in
the potential is strongly reduced when going toward the
drip-line. This reduction is especially pronounced in the
surface region, and consequently one should observe a more
diffuse surface. Since the densities of the $f$ orbitals
are located on the surface, while those of the $p$ orbitals
are contained in the bulk of the nucleus, the effect of
reduction of the spin-orbit interaction will be much
stronger for the 1f$_{7/2}$ - 1f$_{5/2}$ doublet.

In Fig. 6 we display the neutron densities for the N=20
isotones, calculated with the NL3 effective force. The
figure corresponds to the central panel in Fig. 5. In the
insert of Fig. 6 we include the corresponding values for
the surface thickness and diffuseness parameter. The
surface thickness $s$ is defined to be the change in radius
required to reduce $\rho (r) / \rho_0$ from 0.9 to 0.1
($\rho_0$ is the density in the center of the nucleus). The
diffuseness parameter $\alpha$ is determined by fitting the
neutron density profiles to the Fermi distribution
\begin{equation}
\rho (r) =  {\rho_0} \left (1 + exp({{r - R_0}\over 
\alpha})\right)^{-1} ,
\end{equation}
where $R_0$ is the half-density radius. In going away from
the drip line, from $^{28}$O to $^{32}$Mg, the surface
thickness decreases from 2.8 fm to 2.33 fm, and the
diffuseness parameter decreases from 0.64 fm to 0.55 fm.
We have shown the vector densities. However, the scalar 
densities display the same radial dependence. This means
that the scalar potential $S$ and the vector potential
$V$ behave in the same way in the surface region, i.e. 
both potentials display the same diffuseness.

These large changes in surface properties reflect the
reduction of the spin-orbit term of the potential.
In relativistic mean-field approximation, the spin-orbit
potential originates from the addition of two large fields:
the field of the vector mesons (short range repulsion), and
the scalar field of the sigma meson (intermediate
attraction). In the first order approximation, and
assuming spherical symmetry, the spin orbit term can be
written as
\begin{equation}
\label{so1}
V_{s.o.} = {1 \over r} {\partial \over \partial r} V_{ls}(r), 
\end{equation} 
where $V_{ls}$ is the spin-orbit
potential~\cite{Rin.96,KR2.91}
\begin{equation}
\label{so2}
V_{ls} = {m \over m_{eff}} (V-S).
\end{equation}
V and S denote the repulsive vector and the attractive
scalar potentials, respectively.  $m_{eff}$ is the
effective mass
\begin{equation}
\label{so3}
m_{eff} = m - {1 \over 2} (V-S).
\end{equation}
Using the vector and scalar potentials from the NL3
self-consistent ground-state solutions, we have computed
from~(\ref{so1}) - (\ref{so3}) the spin-orbit terms for the
N=20 isotones.  They are displayed in Fig. 7 as function of
the radial distance from the center of the nucleus. The
magnitude of the spin-orbit term $V_{s.o.}$ increases as we
add more protons, i.e. as we move away from the drip-line.
In Fig. 8 we also plot the proton and neutron $rms$ radii
for the N=20 isotones, calculated with all three parameter
sets. The results are very similar, and show how the
neutron radii decrease in going toward the stability line.
At the same time, of course, the proton radii increase.

Finally, in Fig. 9 we compare the neutron $rms$ radii for
Na isotopes with recent experimental data from
Ref.~\cite{Suz.95}.  Theoretical values correspond to
ground-state densities calculated with the NL3 effective
interaction, and the D1S parameters for the Gogny
interaction in the pairing channel.  The blocking procedure
has been used both for protons and neutrons. With the
possible exceptions of N=11, the theoretical values nicely
reproduce the experimental data, much better than the 
RMF parameter set that has been used in Ref.~\cite{Suz.95}. The
neutron $rms$ radii display a uniform increase. Unlike in
the case of Ne, no sudden change in the radii is observed.
Instead of the neutron halo, what we observe in Na isotopes
is more probably the formation of the neutron skin. In
fact, in the insert of Fig. 9 we display the calculated and
experimental differences between neutron and proton radii. The 
smooth increase of $r_n - r_p$ after N=11 presents evidence for a
large neutron skin.  Of course, this does not prevent the
formation of the halo for heavier Na isotopes.  The
discrepancy of the calculated and measured $r_n$ for N=11
probably comes from the fact that for this isotope the
proton and neutron numbers are identical, and therefore
both the odd proton and odd neutron occupy the $s-d$
orbitals with the same probabilities.  One should expect a
short range interaction, possibly proton-neutron pairing.
Such an interaction would lower the total energy, and the
resulting radius should be smaller. An explicit
proton-neutron interaction is not included in the
theoretical model, and therefore the calculated neutron
radius is larger than the experimental value.  
%
\section {Conclusions}

In the present work we have investigated properties of
light nuclei with large neutron excess. Neutron-rich nuclei
between Carbon and Magnesium have recently become available
in experiments with radioactive nuclear beams, and
therefore it is very important to study the properties that
various theoretical models predict for this region. We have
calculated ground state properties using the Relativistic
Hartree Bogoliubov theory.  Based on the relativistic
mean-field model on one side, and the HFB theory on the
other, RHB provides a unified description of mean-field and
pairing correlations. It describes the nucleus as a
relativistic system of baryons and fermions.  Pairing
correlations are described by finite range two-body forces.
Very little is known about the $pp$ component of the
effective interaction in relativistic models, and therefore
we have used a finite range interaction of Gogny-type.
Finite element methods have been used in the coordinate
space discretization of the coupled system of
Dirac-Hartree-Bogoliubov integro-differential eigenvalue
equations, and Klein-Gordon equations for the meson fields.
Coordinate space solutions are essential for a correct
description of particle-hole and pair excitations into the
continuum. For the mean-field Lagrangian we have used the
well established effective interactions NL1, NL3 and NL-SH,
and the D1S parameter set for the finite range Gogny type
interaction in the pairing channel. In particular, we have
investigated the location of the neutron drip-line as
function of the number of protons.

Model calculations have shown that the triplet of
single-particle states near the neutron Fermi level:
1f$_{7/2}$, 2p$_{3/2}$ and  2p$_{1/2}$, and the neutron
pairing interaction determine the location of the neutron
drip-line, the formation of the neutron skin, or eventually
of the neutron halo.  For C, N, O and F the triplet is
still high in the continuum at N=20, and the pairing
interaction is to weak to promote pairs of neutrons into
these levels.  All mean-field effective interactions
predict similar results, and the neutron drip is found at
N=18 or N=20.  For Ne, Na, and Mg the states 1f$_{7/2}$,
2p$_{3/2}$ and  2p$_{1/2}$ are much lower in energy, and
for N$\geq$ 20 the neutrons populate these levels. The
neutron drip can change by as much as twelve neutrons. The
model predicts the formation of neutron skin, and
eventually neutron halo in Ne and Na. This is due to the
fact that the triplet of states is almost degenerate in
energy for N$\geq$ 20.  For Mg the 1f$_{7/2}$ lies deeper
and neutrons above the $s-d$ shell will exclusively
populate this level, resulting in a deformation of the mean
field.  When we plot the neutron single-particle levels as
function of the number of protons, it turns out that the
binding of neutrons is determined by the 1f$_{7/2}$ orbital
which is found to be parallel to the slope of the Fermi
level.  The reduction of the spin-orbit splitting
1f$_{7/2}$ - 1f$_{5/2}$ close to the neutron-drip line has
been related to the isospin dependence of the spin-orbit
interaction.  Density distributions have been analyzed, and
the resulting surface thickness and diffuseness parameter
reflect the reduction of the spin-orbit term of the
effective potential in the surface region of neutron-rich
nuclei. Model calculations reproduce the recent
experimental data on the neutron $rms$ radii for Na
isotopes.

The results of the present work clearly show that a correct
theoretical description of nuclei with large neutron excess
necessitates the following essential ingredients:

\begin{itemize}

\item[i)] self-consistent solution for the
mean-field in order to obtain the correct increase of
the surface diffuseness, 

\item[ii)] the correct isospin dependence of the spin-orbit
term in the mean-field potential. At present only relativistic
models provide reliable results for the
characteristic spacings of single
particle levels at the Fermi surface. 

\item[iii)] pairing correlations describe the scattering
of nucleon pairs for states close to the Fermi surface. 
Pairing leads to the partial
occupation of single particle levels with low orbital
angular momentum.

\item[iv)] pairing correlations in the continuum should
be described by RHB in coordinate space, with
finite-range pairing interaction. In this procedure the 
scattering of nucleon pairs does not lead to 
unbound configurations.

\end{itemize}

Thus Relativistic Hartree-Bogoliubov theory in coordinate
space provides an appropriate framework in which properties
of exotic nuclei on both sides of the valley of
$\beta$-stability can be studied. However, to calculate
ground-state properties of heavier nuclei beyond Na or Mg,
deformations of the mean-field have to be taken into
account. The inclusion of deformation in the coordinate
space formulation of RHB with finite range pairing still
presents considerable difficulties. 
It is also clear that the details of the scattering
processes around the Fermi surface, and the partial
occupation of levels with low orbital angular momentum in
the continuum, are determined by the strength and functional 
dependence of pairing correlations. As long as a
fully microscopic derivation of an effective pairing
in nuclei is not available, it is justified
to use a phenomenological interaction, as for example the Gogny
force. Another open problem is  
whether one needs a density dependent pairing interaction.



\newpage
\leftline{\Large {\bf Figure Captions}}

\begin{description}
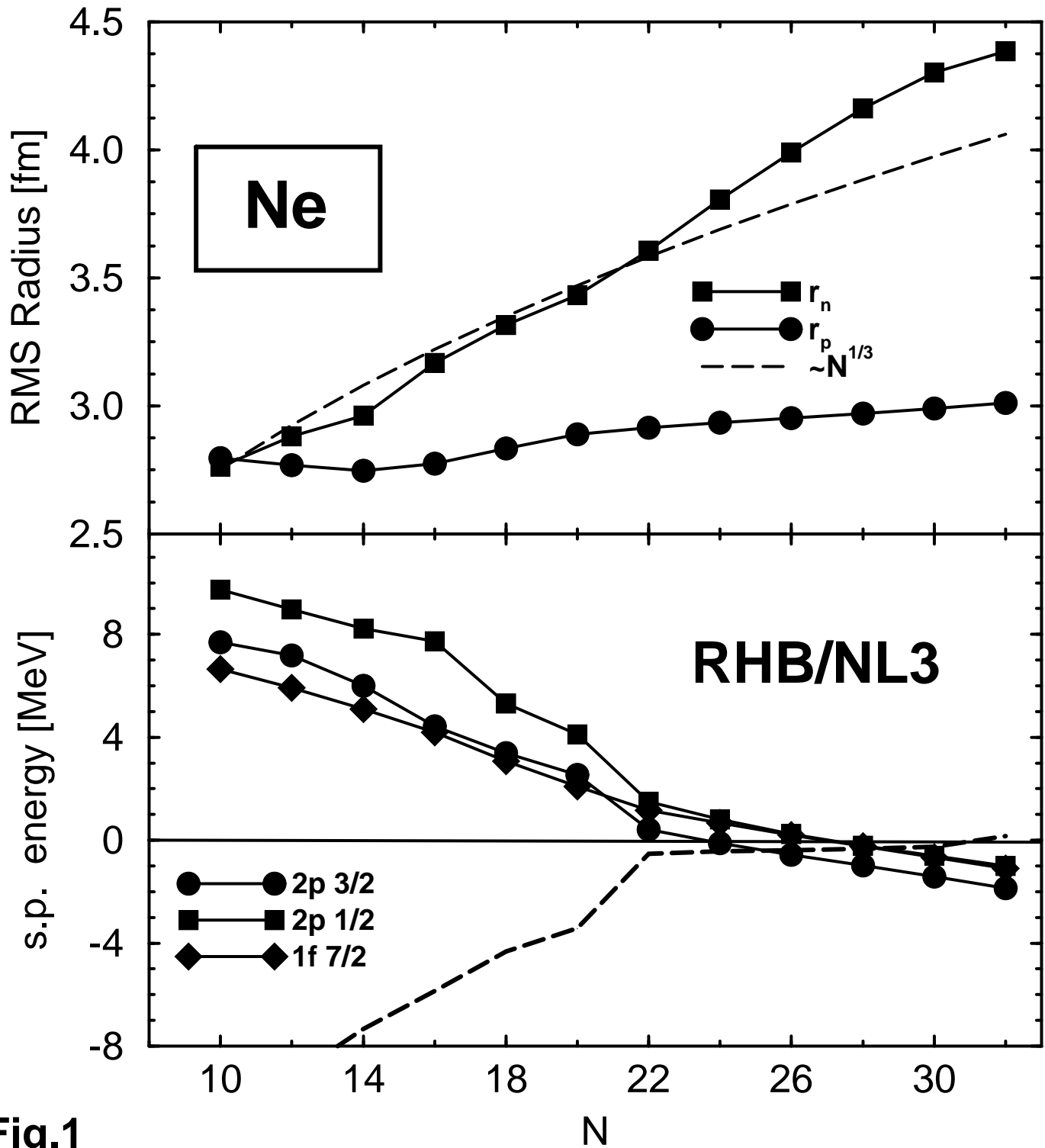
\item{\bf Fig.1} Calculated proton and neutron 
$rms$ radii for Ne isotopes (a), and the 1f-2p single-particle neutron
levels in the canonical basis (b).

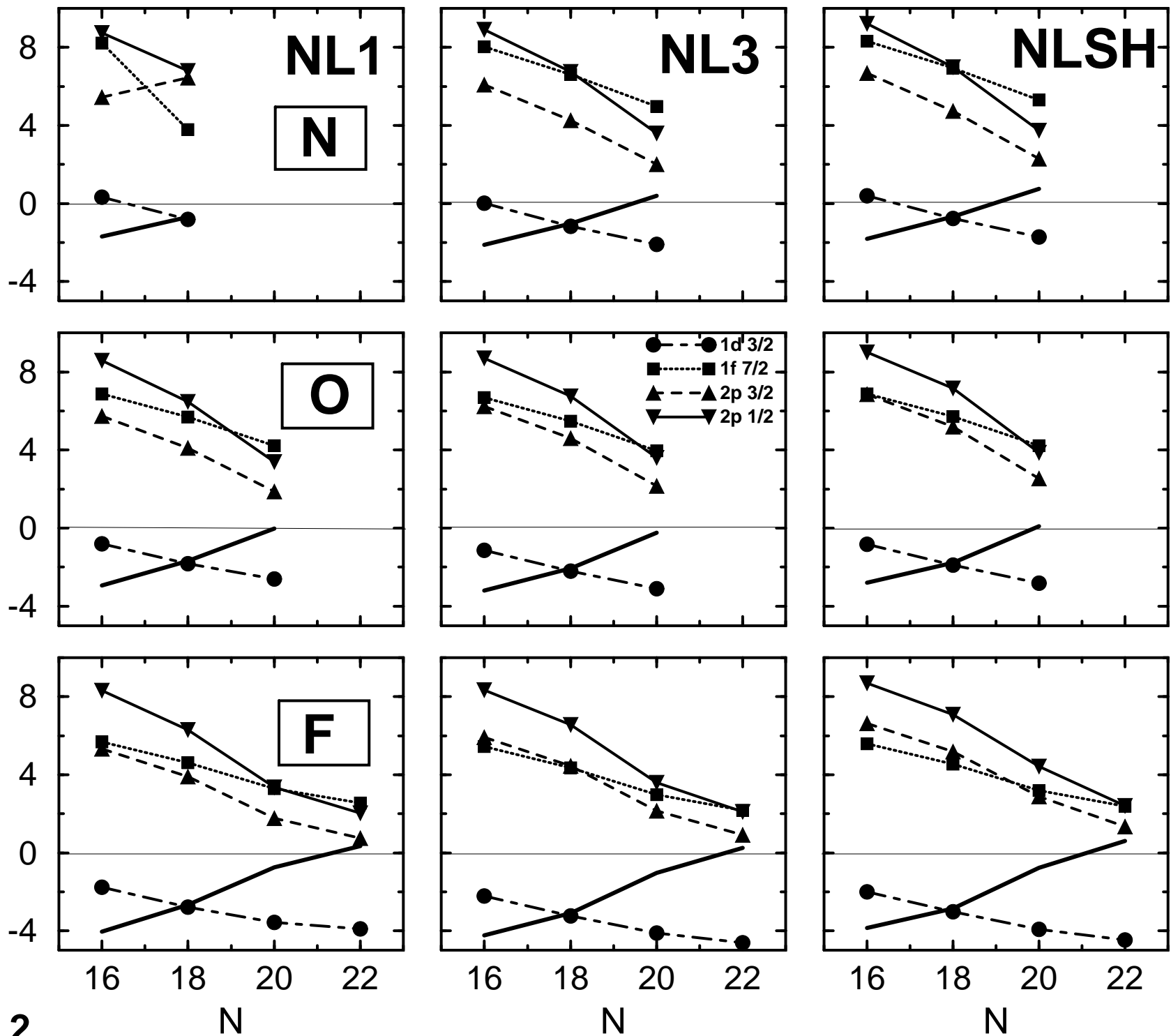
\item{\bf Fig.2} The neutron single-particle levels
1d$_{3/2}$, 1f$_{7/2}$, 2p$_{3/2}$ and 2p$_{1/2}$
for N, O, and F isotopes. Solid lines denote the 
neutron Fermi level. The energies correspond to 
ground-state solutions calculated with the 
NL1, NL3 and NL-SH effective forces of the mean-field 
Lagrangian. The parameter set D1S is used for
the finite range Gogny-type interaction in the 
pairing channel.

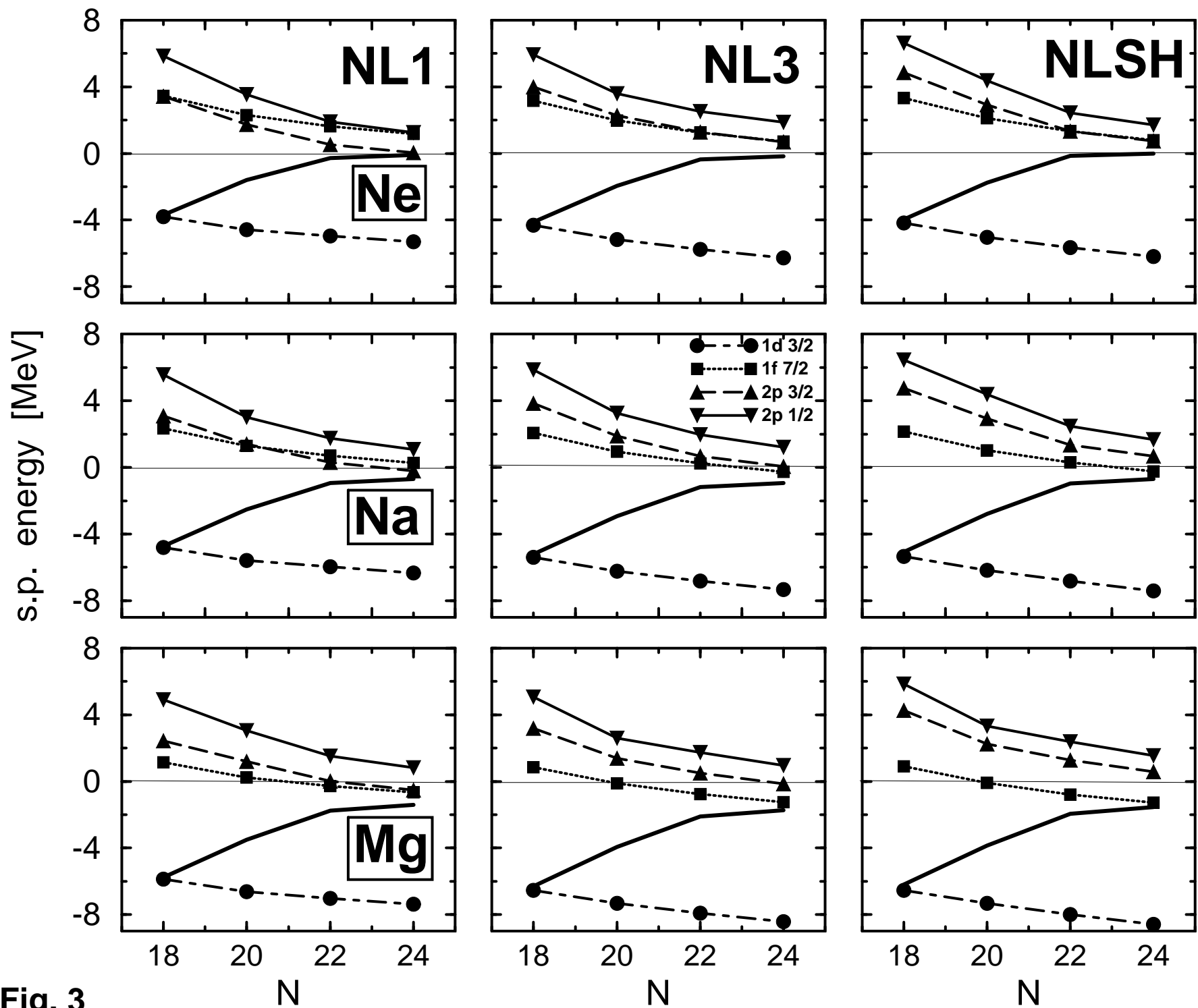
\item{\bf Fig.3} Same as in Fig. 2, but for the 
isotopes of Ne, Na and Mg.

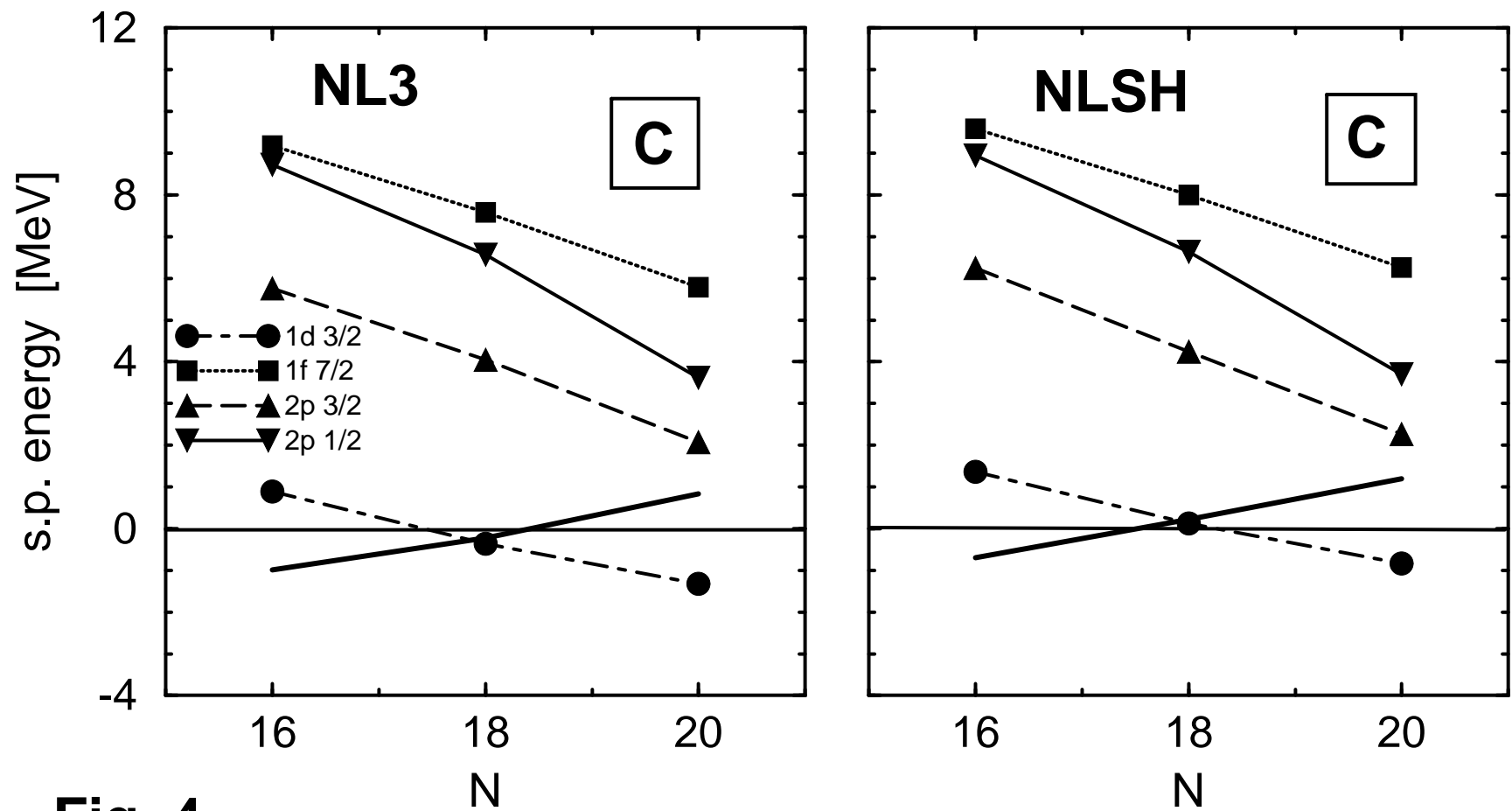
\item{\bf Fig.4} Same as in Fig. 2, but for C isotopes 
and the NL3 and NL-SH effective interactions.
 
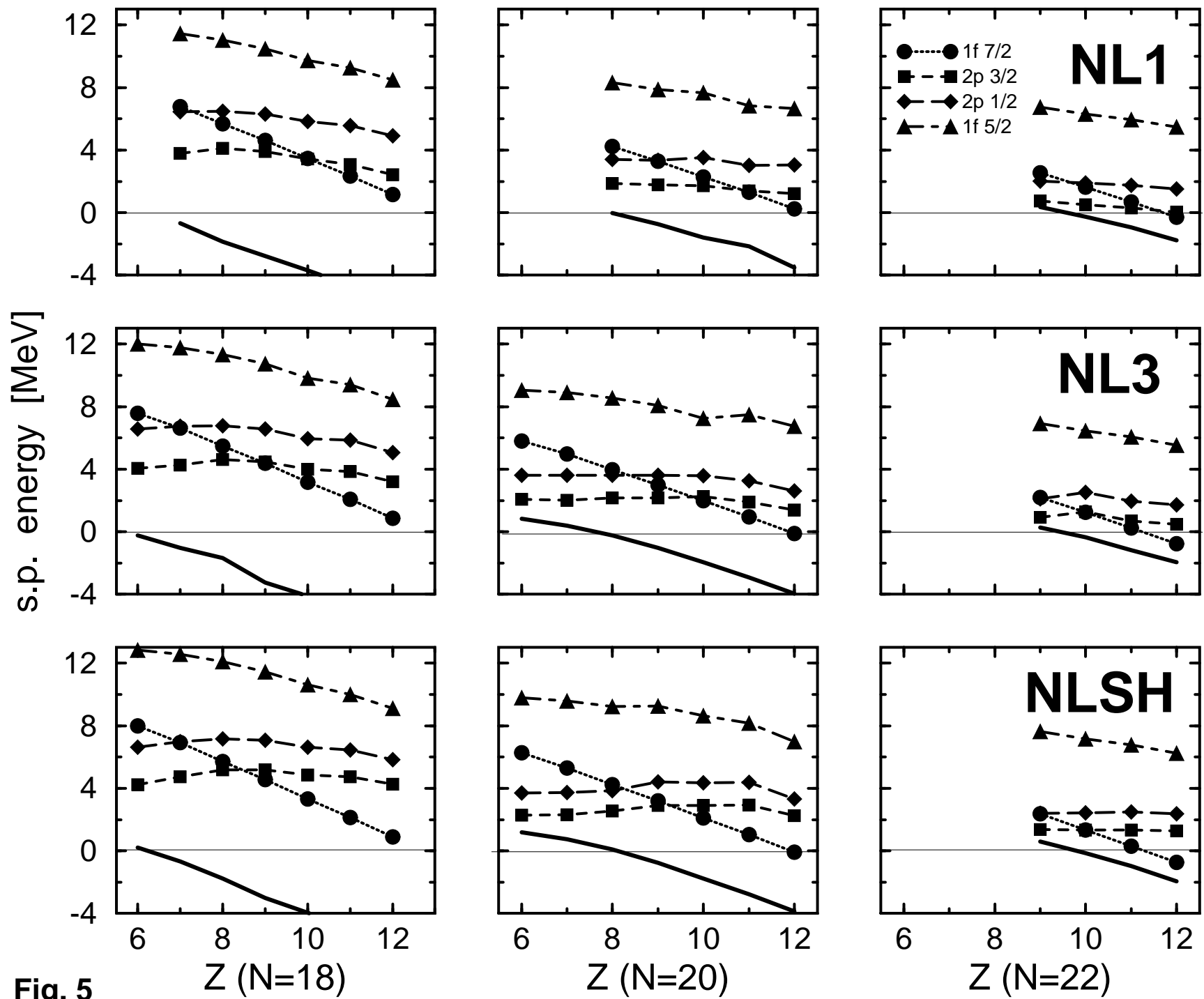
\item{\bf Fig.5} The neutron single-particle levels
2p$_{3/2}$, 2p$_{1/2}$, 1f$_{7/2}$, and 1f$_{5/2}$ 
for the N=18, N=20 and N=22 isotones, as function
of the proton number. Solid lines denote the
neutron Fermi level. The first, second and third row
correspond to calculations performed with the 
NL1, NL3 and NL-SH effective forces, respectively.

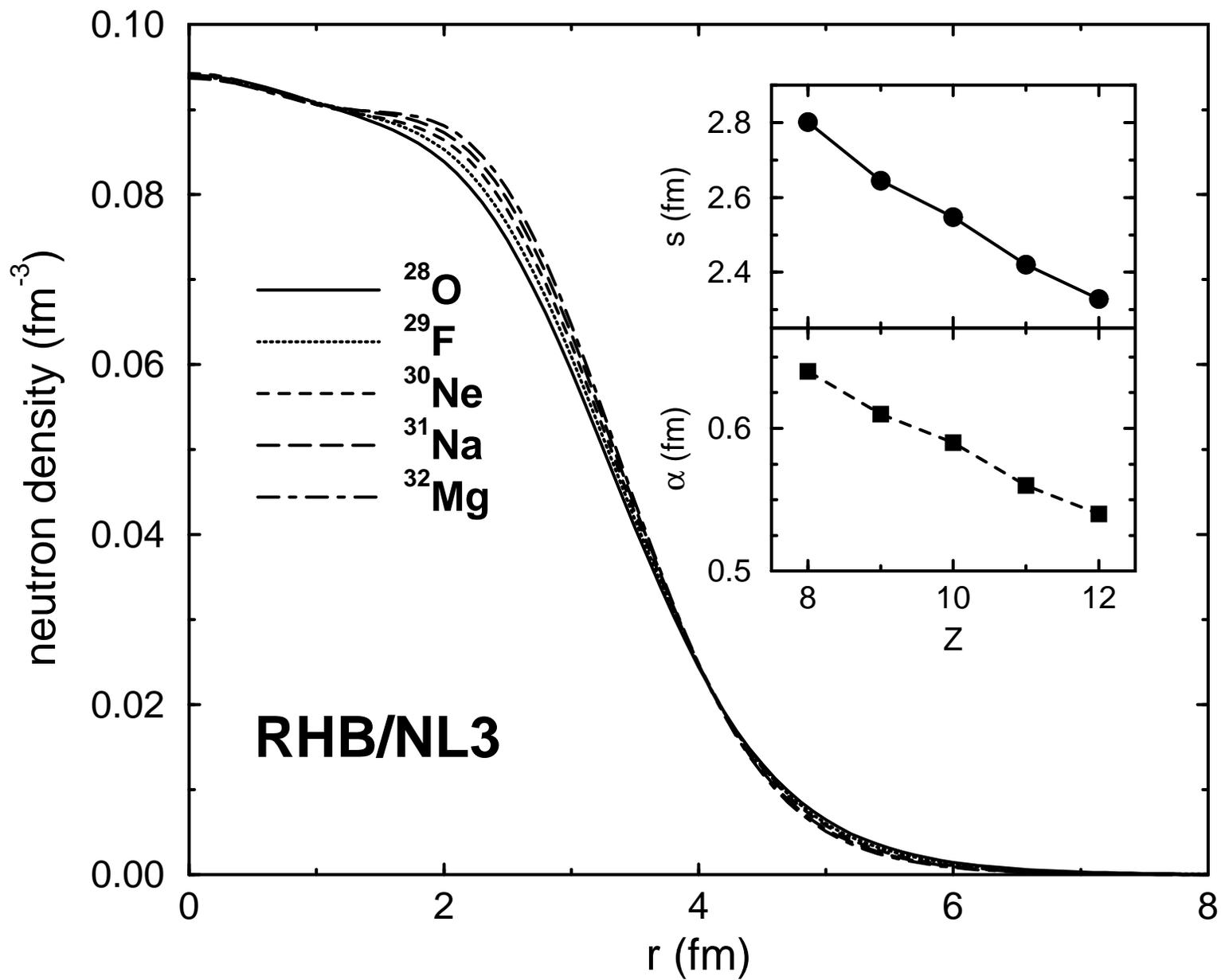
\item{\bf Fig.6} Neutron density distributions for 
the N=20 isotones, calculated with the NL3
effective force. In the insert the corresponding
values for the surface thickness and diffuseness parameter 
are included.

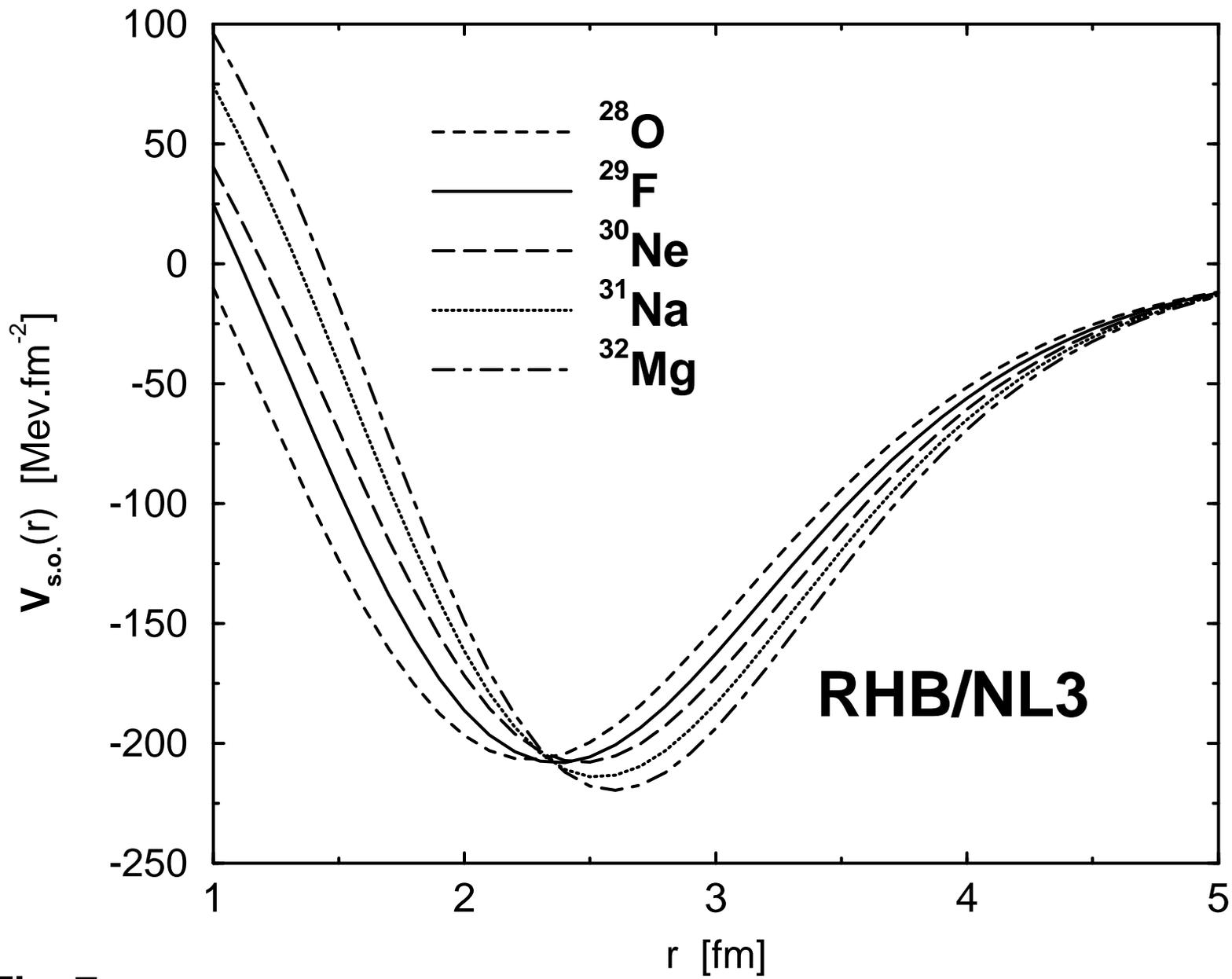
\item{\bf Fig.7} Radial dependence of the spin-orbit term
in self-consistent solutions for the ground-states of
the N=20 isotones, calculated with the NL3 effective force.

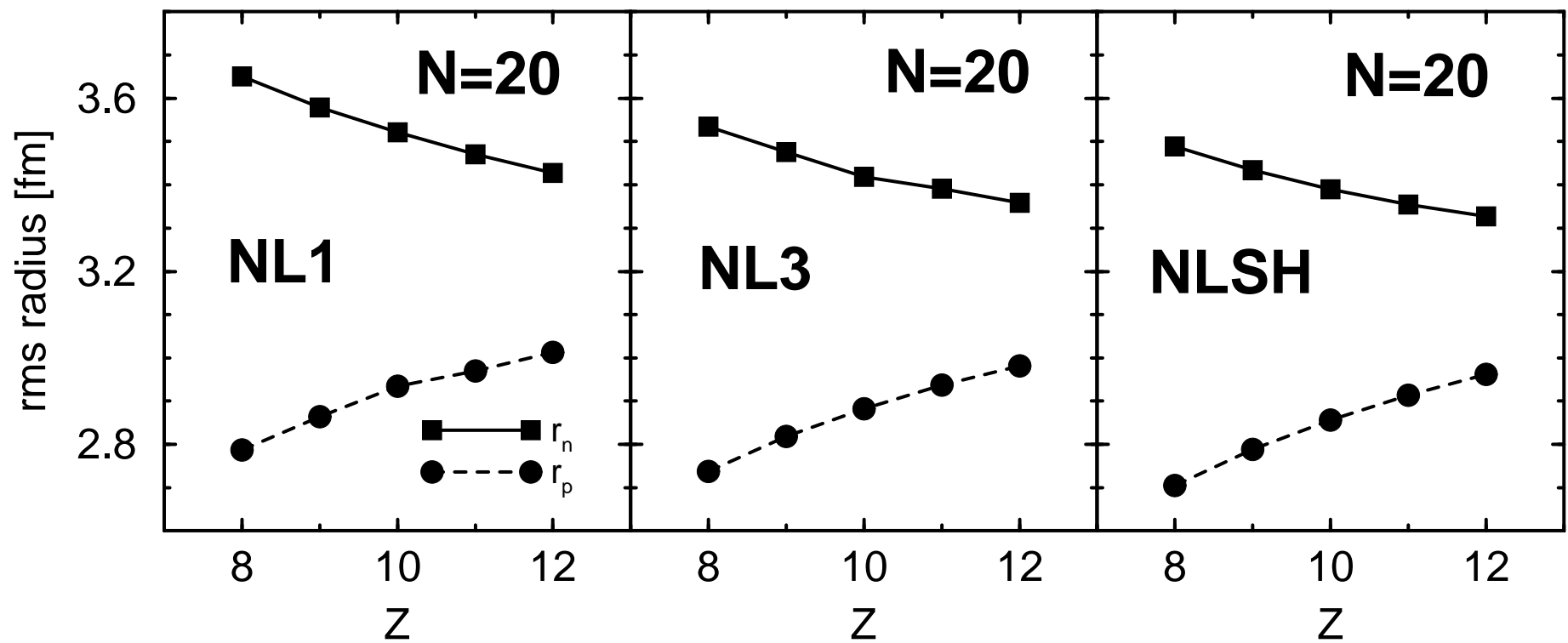
\item{\bf Fig.8} Proton and neutron $rms$ radii for the
N=20 isotones. The panels correspond to results
obtained with the NL1, NL3 and NL-SH effective forces, respectively.

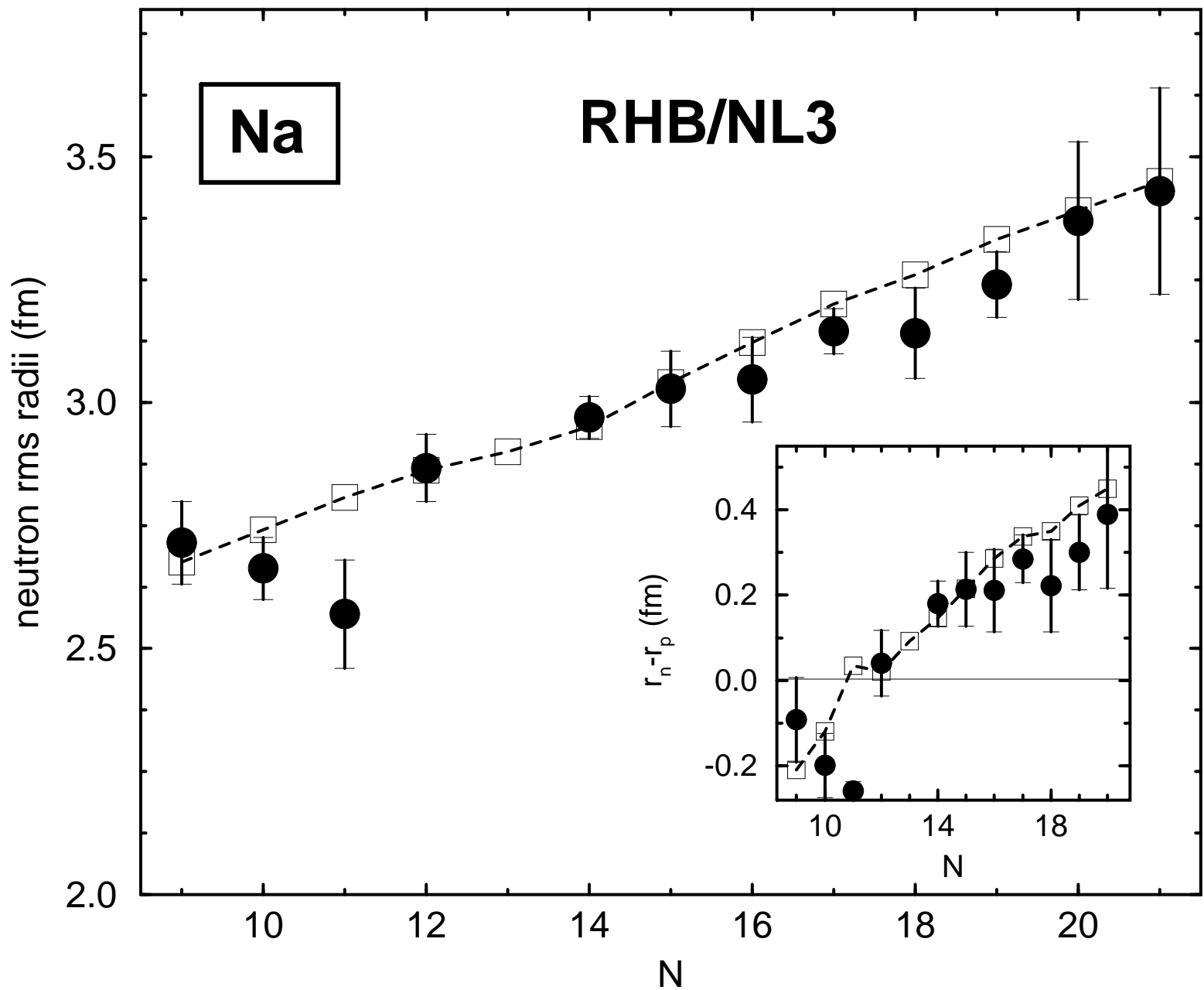
\item{\bf Fig.9} Comparison of the neutron $rms$ radii for Na
isotopes with experimental data from Ref.~\cite{Suz.95}.
Theoretical values (squares) correspond to ground-state densities
calculated with the NL3 effective interaction. The calculated
differences between neutron and proton radii are 
displayed in the insert.

\end{description}
\end{document}